\documentclass[epj]{svjour}

\usepackage{hyperref} 

\usepackage{graphicx,psfrag}
\newcommand{\figfont}{\normalsize}

\usepackage{amsmath}
\usepackage{amssymb}
\usepackage{bbm}

\newcommand{\sfrac}[2]{{#1}/{#2}}

\newcommand{\bk}{\vec{k}}
\newcommand{\bq}{\vec{q}}
\newcommand{\br}{\vec{r}}
\newcommand{\bkappa}{\boldsymbol{\kappa}}
\newcommand{\brho}{\boldsymbol{\rho}}

\newcommand{\ells}{\ell_\text{s}}
\newcommand{\boltz}[1]{{#1}_\text{B}}
\newcommand{\ellB}{\boltz{\ell}}
\newcommand{\tauB}{\boltz{\tau}}
\newcommand{\DB}{\boltz{D}}
\newcommand{\FB}{\boltz{F}}
\newcommand{\WL}[1]{{#1}_{\text{WL}}}
\newcommand{\DWL}{\WL{D}}

\newcommand{\loc}[1]{{#1}_\text{loc}}
\newcommand{\xiloc}{\loc{\xi}}
\newcommand{\tauloc}{\loc{\tau}}
\newcommand{\tauL}{\tau_L}
\newcommand{\tauphi}{\tau_\phi}
\newcommand{\ellphi}{\ell_\phi}
\newcommand{\zetaopt}{\zeta_\text{opt}}
\newcommand{\zetagauss}{\zeta_\text{gauss}}

\newcommand{\cut}[1]{{#1}_\text{c}}
\newcommand{\kc}{\cut{k}}

\newcommand{\av}[1]{\overline{#1}}
\newcommand{\mv}[1]{\left\langle#1\right\rangle}

\newcommand{\eref}[1]{(\ref{#1})}
\newcommand{\fref}[1]{Fig.\,\ref{#1}}

\newcommand{\rmd}{\mathrm{d}}
\newcommand{\dkint}[1]{\int \frac{\rmd\vec{#1}}{(2\pi)^d}\,}

\begin{document}
\title{Quantum Diffusion of Matter Waves in 2D Speckle Potentials}

\author{
C. Miniatura\inst{1,4,5}
\and R.C. Kuhn\inst{2,4} 
\and D. Delande\inst{3}
\and C.A. M\"uller\inst{2}
}
\institute{Institut Non Lin{\'e}aire de Nice, UMR 6618, Universit\'e de Nice Sophia, CNRS; 1361 route
des Lucioles, F-06560 Valbonne, France 
\and 
Physikalisches Institut, Universit\"at Bayreuth,
D-95440 Bayreuth, Germany
\and 
Laboratoire Kastler-Brossel, 
Universit\'e Pierre et Marie Curie-Paris 6, ENS, CNRS;  4 Place Jussieu, F-75005 Paris, France
\and 
Centre for Quantum Technologies, National University of Singapore,
3 Science Drive 2, Singapore 117543, Singapore
\and
IPAL, CNRS, I2R, 1 Fusionopolis Way, Singapore 138632, Singapore
}
\mail{C. Miniatura, \\
\email{christian.miniatura@inln.cnrs.fr}}

\date{\today}

\abstract{
This paper investigates quantum diffusion of
matter waves in two-dimensional random potentials, focussing on 
expanding Bose-Einstein condensates in spatially
correlated optical speckle potentials. 
Special care is taken to describe the effect of dephasing, finite
system size, and an initial momentum distribution.
We derive general expressions for the interference-renormalized diffusion constant, the
disorder-averaged probability density distribution, the variance
of the expanding atomic cloud, and the localized fraction of atoms.
These quantities are studied in detail for the special case of 
an inverted-parabola momentum distribution as obtained from an
expanding condensate in the Thomas-Fermi regime. Lastly, we derive
quantitative 
criteria for the unambiguous observation of localization effects
in a possible 2D experiment.
}

\maketitle

\section{Introduction}

In recent years an increasing number of theoretical and
experimental studies discussed the transport of ultra-cold atoms and
Bose-Einstein condensates in the presence of disorder
\cite{Lewenstein2007} (and references therein). One emblematic phenomenon
in the field is the celebrated Anderson localization phenomenon, a
disorder-induced metal-insulator transition observed in the
absence of inter-particle interactions
\cite{Anderson1958,Tiggelen1999,Kramer1993}. 
Recently two
papers have reported experimental evidence for  
exponential spatial localization of
matter waves using dilute Bose-Einstein condensates 
in a 1D optical speckle potential \cite{Billy2008} and in a quasi-periodic optical
lattice \cite{Roati2008}. 
These results call for investigations in higher
dimensions as it is known that transport in disordered 1D and 2D
potentials always occurs in the localized regime whereas in 3D 
there is a transition \cite{Abrahams1979}. 

In two preceding articles \cite{Kuhn2005,Kuhn2007} 
we presented a theoretical description for the quantum
diffusion of monochromatic non-interacting ultra-cold atoms in 
2D and 3D disordered optical potentials. 
In real experiments, a crucial question in this
context is the influence of a finite
initial momentum distribution of the atoms, together with 
the limits imposed by dephasing processes, like
spontaneous emission, and boundaries. These questions
are addressed in the present paper, focussing on the case of 2D optical speckle
potentials. Coherent diffusive transport of non-interacting matter
waves is studied for any 
initial  
Wigner phase-space distribution within the framework of the
self-consistent theory of localization \cite{Kuhn2007,Vollhardt1980,Shapiro1982}. Within
this formalism, we then
analyze the influence of dephasing processes and of boundaries. The
expected diffusion constant is calculated, both for
the momentum distribution of an interaction-driven expansion of a Bose-Einstein
condensate \cite{Sanchez-Palencia2007} and a Gaussian momentum 
distribution \cite{Shapiro2007}. Finally, we derive criteria for the
experimental observation 
of Anderson localization in 2D correlated
speckle potentials. 

The paper is structured as follows: we first recall in section \ref{principal} some basics about
propagation in disordered systems and calculate the expected probability
density distribution and the variance of the expanding cloud. 
As a first application, we discuss the case of
phase-incoherent, classical diffusion in section \ref{sec:boltzmann}. 
Section \ref{weakloc}  considers coherent
corrections to transport, viz., weak and strong localization. 
We calculate the diffusion constant self-consistently at
finite frequency. In the stationary limit, this leads to the shape of the localized density
distribution and the relevant localization length.
In section \ref{wldephasing.sec}, we incorporate the effect of
dephasing by spontaneous emission into the
formalism, and derive criteria for the effective threshold separating
the diffusive from the (almost) localized regime, where only very slow
residual diffusion occurs. 
Sections \ref{size.sec} and \ref{exp.sec} are devoted to the implications of finite system
size and experimental realizability.

\section{Transport in disordered systems}
\label{principal}

In this paper we describe the dynamics of 
non-interacting 
cold atomic gases evolving in  
static two-dimensional random potentials $V(\vec{r})$.  
If interaction plays no role, a one-body 
description of the atomic dynamics is justified. 
The quantum dynamics is then
described by the Hamiltonian
\begin{equation} \label{eq:hamil}
H=\frac{p^2}{2m}+V(\br),
\end{equation}
where $V(\br)$ 
describes the potential
fluctuations around its average value, which we choose to be the origin
of energies. 
In a single realization of the disorder, 
an initial plane wave $\bk$ 
with kinetic energy $E_k=\hbar^2k^2/(2m)$
will be randomly
scattered by the potential fluctuations to all other
accessible states. As a consequence, only statistical quantities
obtained by configuration average over the disorder realizations can
reveal generic transport properties.  
In the following, configuration averages will be denoted by 
$\av{(\ldots)}$, and as a first property of the disorder potential, $\av{V(\br)}=0$. 

\subsection{Spatial potential correlations}

Experimentally, random potentials $V(\br)$ can be realized as optical speckle potentials generated by
monochromatic illumination and imaging of an appropriate diffusing
plate \cite{Clement2006}. 
In the regime of parameters that we will explore, the pair 
correlation $\av{V(\br+\br') V(\br')}$ is the only 
ingredient one needs to calculate relevant microscopic quantities characterizing the atomic dynamics
\cite{Kuhn2007}.  The potential fluctuations are then
characterized by  the pair correlation
\begin{equation}\label{eq:corr}
\av{V(\br+\br') V(\br')}=V^2 C(r/\zeta)
\end{equation}
that depends only on the modulus $r=|\br|$ when translation and rotation invariance is
restored on average. 
The spatial correlation function $C(\rho=r/\zeta)$ 
decays from $C(0)=1$ to zero over a characteristic
spatial scale $\zeta$. 
In the case of an optical speckle potential this correlation length $\zetaopt =
(\alpha k_L)^{-1}$ is set by the wavelength $\lambda_L=2\pi
/k_L$ of the monochromatic laser field and
by the numerical aperture $\alpha$ of the imaging system (in current
experiments $\alpha \sim 0.3$).

A 2D optical speckle potential created from a circular diffusive
plate is described by \cite{Kuhn2007,Goodman1975} 
\begin{equation}
C_\text{opt}(\rho)=(\sfrac{2J_1(\rho)}{\rho})^2\label{eq:cor2D}
\end{equation} 
where $J_1(\rho)$ is the Bessel function of order 1. 
Its 2D Fourier transform 
 $P (\bkappa)=\int \rmd\brho\,e^{-i\bkappa\cdot\brho}
C(\rho)$, also known as the fluctuation power spectrum,  
is given by  
\begin{equation}\label{cor:Popt}
P_{\text{opt}}(\bkappa) = 8
\left[\arccos\tfrac{\kappa}{2}-\tfrac{\kappa}{2}\sqrt{1-(\tfrac{\kappa}{2})^2}\,\right]
\, \Theta(2-\kappa). 
\end{equation}
The slow algebraic 
decay of the real-space correlations (as $\rho^{-3}$) implies that
their Fourier transform is non-analytic at $\kappa =0$: 
$P_{\text{opt}}(\bkappa) = 4\pi - 4 |\bkappa| + O (\kappa^2)$.  
Also the fact that the power spectrum has finite support in
$k$-space has important
consequences for the localization of matter waves in a 1D geometry, 
allowing for the crossover from 
exponential to 
effectively algebraic
localization in position space 
\cite{Billy2008,Sanchez-Palencia2007}. 

Potential correlations are also often described (or approximated) by a
Gaussian correlator, such as in \cite{Hartung2008}, 
because it is easily implemented in the numerics and leads to somewhat
simpler analytical calculations:   
\begin{subequations}
\begin{align} 
C_\text{gauss}(\rho) & = \exp\left(-\rho^2/2\right),  \\ 
P_\text{gauss}(\kappa) & = (2\pi)^{d/2}\exp\left(-\kappa^2/2\right). \label{eq:corG}
\end{align}
\end{subequations}
If one wishes to match the  
small-$\vec{r}$ expansion of the 
2D optical speckle \eref{eq:cor2D} one must choose $\zeta_{\text{gauss}} = \sqrt{2} \zeta_{\text{opt}}$. 
The power spectrum \eref{eq:corG} decays 
as a Gaussian, which 
is often a qualitatively good approximation to the finite support of
the true optical speckle spectrum \eref{cor:Popt}. 

If only the behavior at very low momenta $\kappa =k\zeta \ll 1$  is of interest, 
the detailed structure of the correlation function is unimportant,
and a good and simple approximation is provided by $P(\kappa) \approx
P(0)$ where $P_\text{opt}(0) = 4\pi$ and
$P_\text{gauss}(0)=(2\pi)^{d/2}$.   
This in turn corresponds to an effectively $\delta$-correlated position-space correlator    
\begin{equation} 
\label{eq:cordelta}
\av{V(\br+\br') V(\br')} =V^2 P(0) \zeta^d  \delta(\br)  . 
\end{equation} 

\subsection{Weak disorder regime}
\label{weak.sec}

The correlation length 
$\zeta$ defines a correlation energy 
\begin{equation} 
E_\zeta = \frac{\hbar^2}{m\zeta^2}
\end{equation}
and its associated time scale  $\tau_\zeta = \hbar/E_\zeta =
m\zeta^2/\hbar$. 
The matter wave dynamics is then driven by three different energy scales:
the kinetic energy $E_k$, the strength $V$ of potential fluctuations, and the correlation
energy $E_{\zeta}$.

The impact of potential fluctuations on the atomic dynamics is simply
estimated by the magnitude $\sfrac{V \tau}{\hbar}$ of the random phase
kick experienced by an atom travelling over the distance $\zeta =v\tau $ at
velocity $v=\hbar k/m$ through potential fluctuations of strength
$V$. As shown in~\cite{Kuhn2007}, 
the weak-scattering regime is realized if 
\begin{equation}\label{weakscattering}
\frac{V \tau}{\hbar} = \sqrt{\frac{E_\Delta}{2 E_k }} = \frac{\eta}{k\zeta}  \ll 1.
\end{equation}
Throughout the paper, we will use 
\begin{equation} 
\eta = \frac{V}{E_\zeta}
\end{equation}
as a measure for the potential fluctuation strength.
Condition \eref{weakscattering} 
shows that the weak-disorder regime is realized if the kinetic
energy $E_k$ is sufficiently above the important energy scale 
$E_\Delta = V^2/E_\zeta$.  
In 3D, this is essentially the mobility edge 
separating extended
states with $E_k> E_\Delta$ from localized ones with $E_k<E_\Delta$ \cite{Kuhn2007}. 

Even within the weak scattering regime $E_k\gg E_\Delta$ one may still be
able to probe the low-energy or $\delta$-correlated potential regime 
$E_k\ll E_\zeta$ provided the following energy 
hierarchy holds: $E_\Delta \ll E_k \ll E_\zeta$. This requires
very weak potential 
fluctuations $V \ll E_\zeta$ or equivalently $\eta \ll 1$.

In the weak-scattering regime  
the atoms experience thus many random phase kicks of small amplitude
in the course of time, 
which repeatedly alter their wave function. These
phase kicks are the seeds of a diffusive behavior for the average probability
density that will be calculated in the following.

\subsection{Phase-space dynamics}

The ensemble-averaged atomic dynamics in a $d$-dimensional potential is fully described by the
average Wigner distribution 
\cite{Hillery1984}: 
\begin{equation}\label{wtdef}
W_t(\br,\vec{k}) = \int \rmd\br' \, e^{-i\vec{k}\cdot\br'} \, 
\left\langle \br+\tfrac{\br'}{2}|\av{\varrho(t)}|\br-\tfrac{\br'}{2}\right\rangle
\end{equation}
where $\av{\varrho(t)}=\av{U(t)\,\varrho_0\,U^\dag (t)}$ is the
one-particle average density operator. Here $\varrho_0$ denotes
the initial atomic density operator and $U(t)=\Theta(t)\exp\{-iHt/\hbar\} $ is the forward-time
evolution operator for the Hamiltonian \eref{eq:hamil}, $\Theta$ being the Heaviside step function.  
From this Wigner function one can extract the marginals $p(\br,t)$
(spatial distribution at time $t$) and $\pi(\vec{k},t)$ (momentum distribution at time $t$) according to:
\begin{align}
&p(\br,t)=\dkint{k}  W_t(\br,\vec{k}),
&\pi(\vec{k},t)=\int \rmd\br \ W_t(\br,\vec{k}).
\end{align}
The normalization of these marginals is 
\begin{equation}\label{eq:norma}
\int \frac{\rmd\vec{k}}{(2\pi)^d} \ \pi(\vec{k},t) = \int \rmd\br \ p(\br,t) = 1,
\end{equation} 
which means that the Wigner distribution $W_t$ is normalized  as 
\begin{equation}
\int \frac{\rmd\br \rmd\vec{k}}{(2\pi)^d} \ W_t(\br,\vec{k}) = 1. 
\end{equation} 

Rewriting the integrand of
\eref{wtdef} by separating the initial Wigner distribution $W_0$ gives 
$W_t$ in the form  
\begin{equation}\label{eq:psd}
W_t(\br,\vec{k}) = \int \frac{\rmd\br'\rmd\vec{k}'}{(2\pi)^d} \, 
G_t(\br, \vec{k} ; \br',\vec{k}') \, W_0(\br',\vec{k}')
\end{equation}
with a rather transparent physical meaning: 
starting from the point
$(\br',\vec{k}')$ at time $t=0$, the initial
quasi-probability $W_0$ is propagated in phase space
with $G_t$ to give $W_t$. The phase-space propagation kernel $G_t$
thus encapsulates the
relevant quantum dynamics and reads:
\begin{subequations}
\begin{align}
&G_t = \int  \rmd\br_1 \rmd\br_2 \,
e^{-i(\vec{k}\cdot\br_1-\vec{k}'\cdot\br_2)} \, 
\mathcal{L}_t(\br, \br', \br_1, \br_2) \\
&\mathcal{L}_t  = \av{K_t(\br+\tfrac{\br_1}{2}, \br'+\tfrac{\br_2}{2})
\, 
K_t^\ast(\br-\tfrac{\br_1}{2}, \br'-\tfrac{\br_2}{2})}.
\end{align}
\end{subequations}
Here the quantum propagator $K_t(\br_1,\br_2) = \langle \br_1 | U(t)|\br_2
\rangle$ is the probability
amplitude to end up at point $\br_1$ at time $t$ when starting at point
$\br_2$ at time $t=0$. In Feynman's path description, 
$K_t$ is expressed as a sum over all possible
paths connecting $\br_2$ and $\br_1$. This assures that
$G_t$ indeed contains the non-local quantum
interference that affects the density dynamics 
in phase space. 
It can be readily checked that $G_t$ satisfies:
\begin{subequations}
\begin{align}
&\lim_{t\to0} G_t(\br, \vec{k} ; \br',\vec{k}') = (2\pi)^d \, \delta(\br-\br') \, \delta(\vec{k}-\vec{k}')\\
&\int \frac{\rmd\br \rmd\vec{k}}{(2\pi)^d} \ G_t(\br, \vec{k} ; \br',\vec{k}')  = 1,
\end{align}
\end{subequations} 
where the latter merely expresses, as it should, conservation of the total density.

\subsection{Disorder-averaged probability density}

Using the previous results, it is easy to show that the average
number density distribution is given by 
\begin{equation}\label{eq:wip}
p(\br,t)=\int\frac{\rmd\br' \rmd\vec{k}'}{(2\pi)^d}\;
F_t(\br-\br',\vec{k}')\;W_0(\br',\vec{k}'),
\end{equation}
with a spatial propagation kernel
given by 
\begin{equation}
F_t(\br-\br',\vec{k}') = \dkint{k} G_t(\br,
\vec{k}; \br',\vec{k}'). 
\end{equation}
Taking advantage of the translation invariance, one can develop  
the intensity kernel into its Fourier components at fixed momentum $\bk$,  
\begin{equation}
\label{FourierKernel}
\Phi(\omega, \vec{q}, \vec{k})= \int \rmd\br \rmd t \, e^{i(\omega t - 
\vec{q}\cdot \br)} \, F_t(\br,\vec{k}).  
\end{equation} 
Probability conservation and linear response theory \cite{Kuhn2007}  imply that  
for long times and distances, i.e., small $\omega$ and $q$, the kernel
takes the diffusive form 
\begin{equation}
\label{DiffusionKernel}
\Phi(\omega, \vec{q}, \vec{k}) = \frac{1}{-i\omega + D(\omega, \vec{k}) \, q^2}
\end{equation}
featuring the $\omega$-dependent
diffusion constant $D(\omega, \vec{k})$ that can be calculated
microscopically by quantum transport theory \cite{Kuhn2007,Rammer1998}. 
In this paper, we only consider situations where the disorder average 
restores translational  and rotational symmetry, as
exemplified by the potential correlator  
\eref{eq:corr}. This
assumption implies in particular
that the diffusion constant $D(\omega,\bk)$ must be a scalar and can only depend on $|\vec{k}|=k$. 
Before
turning to the microscopic theory, we first explore which are the observable consequences
of the diffusive form \eref{DiffusionKernel} in transport
experiments with cold atoms.  

\subsection{Variance of the expanding cloud}

The position variance of the expanding cloud of cold
atoms in the speckle potential gives direct access to the
diffusion constant. Indeed, 
the variance of the expanding cloud is defined by $\Delta \br^2 = \mv{\br^2} -\mv{\br}^2$
where $\mv{...}$ denotes the average with respect to $p(\br,t)$.
Using \eref{eq:wip}, \eref{FourierKernel}, and \eref{DiffusionKernel}
one finds:
\begin{equation}\label{variance}
\Delta \br^2(t) = \Delta \br_0^2 - 2d \int \frac{\rmd\omega}{2\pi} \, \frac{\mathcal{D}(\omega)}{(\omega+i0)^2} \ e^{-i\omega t} 
\end{equation}
where $\Delta \br_0^2$ is the
initial variance of the atomic cloud. The notation $\omega+i 0$
indicates that the integration over
frequency has to circumvent the pole at
$\omega=0$ from above in order to describe forward-time
propagation. Here, the effective diffusion constant 
\begin{equation}\label{averageD.eq}
\mathcal{D}(\omega) = 
\mv{D(\omega,k)}_{\pi_0}= 
\dkint{k} \pi_0(\vec{k}) \, D(\omega,k)
\end{equation}
is the $k$-dependent diffusion constant averaged over the initial
momentum distribution $\pi_0(\bk)$.  

\subsection{Interaction-driven momentum distribution}

We wish to calculate the average diffusion constant $\mathcal{D}(\omega)$ using
the momentum distribution of an expanding 
Bose-Einstein condensate (BEC) released from an isotropic harmonic trapping potential with frequency
$\Omega$. Initially, the interacting condensate with chemical
potential $\mu \gg \hbar \Omega $ is
trapped in a Thomas-Fermi parabola with radius $R$ determined by
$\mu = \frac{1}{2} m \Omega^2 R^2$. 
When the harmonic trap is switched
off, the expansion occurs in two steps. For small times $\Omega t\ll1$, 
the repulsive interactions
drive the expansion and the condensate acquires a dynamical phase. For
long times $\Omega t\gg1$, the density drops so much that the expansion becomes free and the
condensate wavefunction takes the form \cite{Kagan1996} 
\begin{equation}
\phi(\br,t)\propto\sqrt{1-z^2} \ \Theta(1-z)\,\exp(i u z^2), 
\end{equation}
with $z =r/(\sqrt{2} R \Omega t)$ and 
$u = 2\mu t/\hbar$. 
Since $\Omega t\gg 1$ and
$\mu /(\hbar \Omega) \gg1$, one can use a stationary-phase
approximation with $u\gg 1$ to calculate the Fourier transform to the momentum
domain, which is stationary: 
\begin{equation}
\label{eq:philo}
\tilde{\phi}(\vec{k}) \propto \sqrt{1-(k\xi)^2}\,\Theta(1-k\xi). 
\end{equation}
Here, $\xi = \hbar /\sqrt{4m\mu}$
 is the healing length defined such that $\mu=\hbar^2/4m\xi^2$.  
Hence, the normalized momentum distribution
$\pi_0(\vec{k})=|\tilde{\phi}(\vec{k})|^2$ is given by 
\begin{equation}\label{eq:pilaur}
\pi_0(\vec{k}) = \frac{(2\pi\xi)^d}{S_d}\frac{d(d+2)}{2} \;(1-(k\xi)^2)\,\Theta(1-k\xi)
\end{equation}
where $S_d$ denotes the surface of the $(d-1)$-dimensional unit sphere ($S_1=2,S_2=2\pi,S_3=4\pi$). 
This distribution has again the form of an inverted parabola with an upper cut-off $
k_{\mathrm{max}} = 1/\xi$ \cite{Sanchez-Palencia2007}.

When the asymptotic regime of free expansion is reached, the speckle potential
is switched on. The distribution $\pi_0(\vec{k})$ then constitutes the
initial momentum distribution for the evolution of the condensate in
the disordered potential. 

\section{Classical transport: the Boltzmann picture}
\label{sec:boltzmann}

In order to put the above results to work, let us first calculate the
dynamics of particles in a disordered potential if all interference
effects can be neglected. This is the realm of classical transport
theory, whose origins date back to the Drude-Boltzmann
theory of metals, more than one century ago 
\cite{Rammer1998,Ashcroft1976,Akkermans2007}, 
and that was later adapted to light propagation through
interstellar atmospheres (radiative transfer theory)
\cite{Chandrasekhar1960}. The basic physical ingredients of this description are
(i) first to assume that any possible interference effects are washed
out under disorder average (random-phase assumption) and (ii) second
to devise a detailed-balance analysis of energy transfer in
phase space (scattering, absorption, sources, etc). This powerful
description leads to a physically very simple
and appealing picture: sufficiently far from the boundaries and in the
hydrodynamic regime of long times and large distances, transport
is described by \textit{diffusion}. 
Note that the description of transport in terms of a Boltzmann
equation as such---and by its diffusive limit at long time and large
distances---is \emph{not} restricted 
to the weak scattering limit established in section \ref{weak.sec}
above. 
However, it is in the weak-scattering regime that 
macroscopic quantities can be easily computed  
from the microscopic ones as will be shown in the following. 

\subsection{Boltzmann transport mean free path}
\label{ellB.sec}

When all interference corrections to the intensity kernel can be
neglected (see section \ref{wldephasing.sec} below for a more detailed derivation) 
one finds a finite diffusion constant in the stationary limit of
frequency $\omega\to 0$ given by \cite{Kuhn2007}
\begin{equation} 
\DB(k)=\frac{\hbar}{md} k \ellB(k)= \frac{v\ellB}{d}=\frac{\ellB^2}{\tauB}
\label{DB}
\end{equation}
Note that this expression formally remains 
valid beyond the weak scattering limit, provided the 
full dispersion relation 
(instead of the free-particle expression $v=\hbar k/m$) is used 
for the average velocity. 
The Boltzmann transport mean-free path $\ellB$ is the
characteristic spatial scale beyond  
which memory of the initial direction is lost. 
It thus identifies with the average step of the random walk induced by
the scattering processes.  
This classical transport mean-free path can be calculated 
both from the microscopic theory \cite{Kuhn2007} and 
from a detailed-balance \textit{Ansatz}, by an angular integral over
the differential cross section of a single scattering event,   
weighted by the scattering anisotropy factor $(1-\cos\theta)$:    
\begin{equation}
\label{lb.general}
\frac{1}{k \ellB} = \frac{\eta^2}{\kappa^{4-d}} 
\int \frac{\rmd\Omega_d}{(2\pi)^{d-1}} (1-\cos\theta)
P\left(2\kappa\sin(\theta /2)\right). 
\end{equation}
Here $\kappa = k\zeta$, $d\Omega_2=d\theta$ in 2D (integration range
from $0$ to  $2\pi$) and $d\Omega_3=2\pi \sin\theta \,d\theta$ 
in 3D (integration range from $0$ to $\pi$). 

In the weak-scattering
regime, the elastic scattering mean-free
path $\ells$ is given by expression \eref{lb.general} with the anisotropy
factor replaced by 1. 
For the correlated potentials considered here (either the speckle or
the Gaussian one), $\ellB$ 
is larger than $\ell_s,$ the two being asymptotically equal in the low-energy limit where scattering is isotropic.

An analytic expression for
$\ellB$ is available 
in the small- and large-momentum limit \cite{Kuhn2007} 
for the 2D optical speckle correlator \eref{eq:cor2D} where
$\zeta=\zetaopt$: 
\begin{subequations}
\begin{align}
&k\ellB = k\ells \approx \frac{(k\zeta)^2}{4\pi\eta^2},   \qquad &k\zeta \ll 1, \label{lowk} \\
&k\ellB = \frac{15}{4} \, (k\zeta)^2 \, k\ells \approx 
\frac{45\pi (k\zeta)^5}{128 \eta^2}, \qquad &k\zeta \gg 1. \label{largek}
\end{align}
\end{subequations}
For the Gaussian correlation \eref{eq:corG} with $\zeta=\zetagauss$, one finds in 2D an
analytical result valid for all momenta $k$, 
\begin{equation} 
\label{lB.gauss}
\frac{1}{k \ellB} = \frac{2\pi \eta^2}{k^2\zeta^2}e^{-k^2\zeta^2} 
\left[ I_0(k^2\zeta^2) - I_1(k^2\zeta^2) \right], 
\end{equation}
where $I_\nu$ is the modified Bessel function. This result has also been derived in \cite{Apalkov2004} in a different context. In the low-energy 
limit $k\zeta\ll1 $ both optical speckle and Gaussian
correlations result in the same expression 
\begin{equation} 
\label{lB.delta}
\frac{1}{k \ellB} = \frac{P(0) \eta^2}{k^2\zeta^2} = \frac{P(0)
E_\Delta}{2E_k}, 
\end{equation}
just as substituting $P\left(2\kappa\sin(\theta /2)\right) \approx P(0)$ in
\eqref{lb.general} would have given. Hence $k\ellB \gg 1$ in the
weak-disorder regime $E_k \gg E_\Delta$.

On the high-momentum side, 
it is important to note that contrary to the 1D case
\cite{Billy2008,Sanchez-Palencia2007}, in 2D the transport mean-free path $\ellB(k)$ is
finite for all $k$-values, even though the potential correlator may
have finite support, because even fast atoms can be deflected ever so
slightly by a smooth random potential.

\subsection{The Diffuson kernel}

With the frequency-independent diffusion constant $\DB(k)$, the intensity propagation kernel is  
easily calculated from \eref{FourierKernel} and \eref{DiffusionKernel} and reads 
\begin{equation}\label{eq:ladder}
\FB(r,k,t) = (4\pi \DB(k) t)^{-\sfrac{d}{2}}\,
\exp[-\sfrac{r^2}{(4\DB(k)t)}].
\end{equation}
This Boltzmann kernel, known as the \textit{diffuson} \cite{Akkermans2007},
obeys the diffusion equation:
\begin{equation}\label{eq:difu}
\partial_t \FB - \DB\nabla^2 \FB
= \delta(\br)\delta(t).
\end{equation}
The diffuson obviously fulfills probability conservation. Interestingly enough, 
it also boils down to $\delta(\br)$ when $t \to 0$ even though its expression 
is in principle only valid at large enough times. 
The range of validity of this  
diffusive description is then expressed as $r\gg \ellB$ and $t\gg
\tauB$ in terms of 
the Boltzmann transport time $\tauB=\ellB^2 / \DB$. 
 With the 
interaction-driven momentum distibution \eref{eq:philo}, 
this will occur for the whole distribution as soon as
\begin{equation}
\frac{4\mu t}{\hbar d} \gg k_{\mathrm{max}}\ellB \quad , \quad
\frac{r}{\xi} \gg k_{\mathrm{max}}\ell_\text{max} 
\end{equation} 
where $\ell_\text{max}=\ellB(k_{\mathrm{max}} = 1/\xi)$.

Since diffusion in phase space occurs with a momentum-dependent diffusion constant, 
the average probability density $p(r,t)$ itself does not obey a diffusion equation. 
But going back to \eref{eq:wip}, we note that for times large enough, the spatial 
width of the diffuson will always be much larger than the spatial width of $W_0$. 
This implies that, in the long-time limit, the probability density is
well approximated by
the momentum-distribution average  
\begin{equation}
p(r,t) \approx  \mv{\FB(r,k,t)}_{\pi_0} . 
\end{equation}

Finally, with the frequency-independent diffusion constant $\DB(k)$, one finds
that the 
variance \eref{variance} of the expanding cloud increases as expected linearly in time, 
\begin{equation}\label{varianceB}
\Delta \br^2(t) = \Delta \br_0^2 + 2d   \mathcal{D}_B t, 
\end{equation}
with the momentum-averaged diffusion constant analogous to \eref{averageD.eq},
\begin{equation} \label{averageDB.eq}
\mathcal{D}_B = \mv{\DB(k)}_{\pi_0}. 
\end{equation}
This average over the inverted-parabola distribution \eref{eq:pilaur}. 
can be easily evaluated numerically for any of the mean-free paths 
\eref{lb.general}-\eref{lB.delta} above. An analytical result is available in the low-energy limit: 
\begin{equation}
\mathcal{D}_B = \frac{\DB(k_\text{max})}{3} = \frac{1}{3} \ \frac{\hbar}{2m} \ \frac{2E_\text{max}} {P(0)E_\Delta }
\end{equation}
where $E_\text{max}=2\mu$ in terms of the initial chemical
potential. This is a good approximation only if the momentum
dispersion is so small  
that $k_{\text{max}}\zeta=\zeta/\xi \ll 1$,  i.e., in the limit where
already the initial condensate is weakly interacting. 

\section{Coherent transport and  
localization}
\label{weakloc} 

The ensemble average over static 
disorder alone cannot wipe out interference effects, and
Boltzmann's description does not adequately describe phase-coherent propagation.  
Indeed, the constructive
interference between amplitudes 
counter-propagating along 
loop-like paths survives
the disorder average and increases the particle's probability to
return to its
starting point. This increased tendency to stay behind translates into
a reduced diffusion constant, an
effect called weak localization (WL) \cite{Rammer1998,Akkermans2007,Akkermans1995}. 
Under suitable conditions, this can even completely inhibit diffusion, a phenomenon known as strong 
(or Anderson) localization (SL). It is rigorously proven that
transport in 1D bulk samples is always suppressed by strong
localization, and on the basis of the single-parameter scaling theory this
remains valid also in 2D,  
whereas
in 3D localization occurs  
only when $k\boltz{\ell} \lesssim1$ (Ioffe-Regel criterion); further
details can be found in the recommendable review \cite{Tiggelen1999}. 

\subsection{``Loop-renormalized'' diffusion constant}

Phase-coherent bulk transport in the weak localization regime $k\ellB \gg 1$ can be
described by a diagrammatic perturbation theory, 
developped by Vollhardt and W\"olfle  \cite{Vollhardt1980,Rammer1998}. 
When weak-localization interference
effects are 
self-consistently taken into account, the Boltzmann diffusion constant
$\DB(k)$ is renormalized into 
an interference-reduced   
diffusion constant $D(\omega, k)$ solving 
\begin{equation}
\frac{1}{D(\omega, k)} = \frac{1}{\DB(k)} + \frac{A_d(k)}{\DB(k)} C(\omega,k)
\end{equation}
where $A_d(k) = (2\pi)^d \hbar/ (m \pi S_d k^{d-2})$. 
This formula features the ``return probability'' $C(\omega,k)$, given by
\begin{equation}\label{returnproba}
C(\omega,k) = \dkint{q}\frac{1}{-i\omega + D(\omega, k) \, q^2}
\end{equation} 
in the diffusive regime. A mean-field like, self-consistent
description is obtained because  
the corrected diffusion constant $D(\omega,k)$ enters the diffusive
propagator under the integral.   
The implicit equation for $D(\omega,k)$ can be conveniently rewritten as
\begin{equation}\label{SCD}
\DB(k) = D(\omega,k) + \frac{\hbar}{\pi m k^{d-2}} \int \frac{q^{d-1}
\rmd q}{q^2 - i\omega/D(\omega,k)}. 
\end{equation}
In our previous articles \cite{Kuhn2005,Kuhn2007}, we have studied directly
the stationary regime at $\omega=0$. Presently, we will explore the
consequences of the finite-frequency approach, which turns out  to be quite powerful. 


The formal integral for the return probability \eref{SCD} may present
divergencies, either infrared (for small $q$) or ultraviolet (for
large $q$), that have to be regularized by suitable cutoffs. Working at finite  
frequency $\omega$ has the advantage that the integral stays finite as
$q \to 0$, so there is no need for introducing a special infrared cutoff.  

However, the return probability in 2D displays a logarithmic
ultraviolet divergence ($q \to \infty$). We regularize it by
introducing an appropriate upper bound $q_\text{max}$, defined as the minimal distance
beyond which interference can play a role. Since the return probability
is evaluated in the diffusive regime, we have chosen this minimal
distance to be the Boltzmann transport mean free path and accordingly
imposed $1/\ellB$ as the upper bound.  
Please note that 
there is no need to dress also this ultraviolet cutoff by localization
corrections: during the short time  required to scatter through small
loops, corresponding to $\omega \approx \tau _B^{-1}$, the
interference corrections will be shown to be negligible 
just below.

\subsection{Weak and strong localization in 2D}
\label{WSL.sec}

Solving now \eref{SCD} for $d=2$, we obtain the functional dependence 
of $X=D(\omega, k)/\DB(k)$ on $\omega$ at fixed $k$ through the
implicit equation \cite{Lobkis2005}: 
\begin{equation}\label{frequencyD}
-i \omega \tauB = \frac{X}{\exp[\pi k \ellB (1-X)] - 1}. 
\end{equation}
In the weakly disordered regime $k \ellB \gg 1$, it is advisable to
rewrite this transcendental equation as  
\begin{equation}\label{Dlogfrequency}
X = 1 - \frac{1}{\pi k \ellB } \log\left[1 -\frac{X}{i \omega \tauB}
\right] , 
\end{equation}
which gives as perturbative solution the celebrated weak-localization
correction \cite{Gorkov1979,Bergmann1984}   
\begin{equation}\label{WL}
\frac{\DWL(\omega,k)}{\DB(k)}  = 1 - \frac{1}{\pi k \ellB } \log\left[1
-\frac{1}{i \omega \tauB}\right].   
\end{equation}
In the frequency range $\omega\tauB\approx 1$ of small diffusive loops, 
this correction is negligible. This means that the diffusion of a particle starts off 
with a classical Boltzmann random walk and is consistent 
with the choice of the unrenormalized ultraviolett cutoff
$q_\text{max}=1/\ellB$ discussed in the previous subsection. 

\begin{figure}
\psfrag{els}{$\ell_s/\zeta$}
\psfrag{elb}{$\dfrac{\ellB}{\zeta}$}
\psfrag{loc}{$\dfrac{\xiloc}{\zeta}$}
\psfrag{kzeta}{$k\zeta$}
\includegraphics[width=0.9\linewidth]{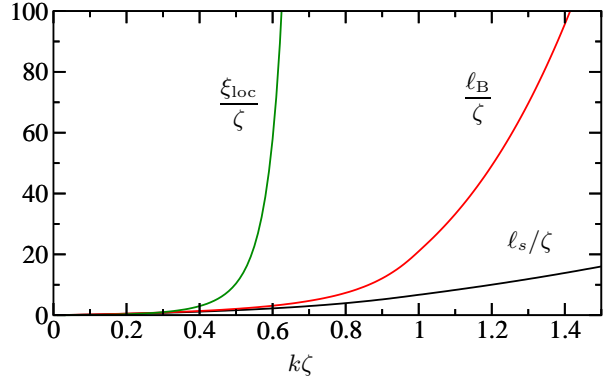}
\caption{Scattering mean free path $\ell_s$, Boltzmann transport mean
free path $\ellB$ [eq.~\eref{lb.general}] 
and localization length $\xiloc$ [eq.~\eref{xiloc}]  as a function of the  atomic wave vector
$k$, all measured in units of the correlation length $\zeta=\zetaopt$ 
of a 2D speckle potential. The disorder strength
is $\eta=V/E_\zeta=0.2$.  
Note the extremely rapid increase
of the localization length, a characteristic feature of 2D localization.}
\label{fig:length}
\end{figure}

At intermediate times $\omega\tauB\ll1$  the weak localization
corrections described by \eref{WL} kick in.    
This expression is valid as  long as the correction is not too large, i.e., for 
$\omega\tauB > (\ellB/\xiloc)^2$ in terms of the new (and important) spatial scale
\begin{equation}
\label{xiloc}
\xiloc(k) = \ellB(k) \sqrt{\exp[\pi k\ellB(k)] -1} \approx \ellB e^{\pi k\ellB/2} 
\end{equation}
whose significance will become clearer in a moment. 
For even smaller frequencies $\omega\tauB \ll (\ellB/\xiloc)^2$ the solution 
of \eref{frequencyD}  crosses over to the asymptotic behaviour   
\begin{equation}
\label{Domegaxi}
D(\omega,k) \approx -i\omega \xiloc^2(k). 
\end{equation}
This linear dependence on frequency with $\lim_{\omega \to 0} D(\omega, k) = 0$  implies 
that the diffusion stops in the long-time limit. The intensity propagation 
kernel \eref{FourierKernel} calculated with the 
asymptotic solution \eref{Domegaxi} is stationary: 
\begin{equation}
F_\infty(r, k) = \int \frac{\rmd\bq}{(2\pi)^2}\, \frac{e^{i\vec{q}\cdot\br}}{1+ q^2\xiloc^2} = \frac{K_0(r/\xiloc)}{2\pi\xiloc^2}
\end{equation}
where $K_0$ is a modified Bessel function~\cite{Abramowitz1972}. 
The asymptotics for large distances 
$K_0(x) = \sqrt{\pi/2x}\, \mathrm{e}^{-x}$
shows an exponential decay, and we see that $\xiloc(k)$ is the localization length.

\begin{figure}
\psfrag{Position}{Position ($\mu$m)}
\psfrag{Probability density}{Probability density}
\includegraphics[width=0.9\linewidth]{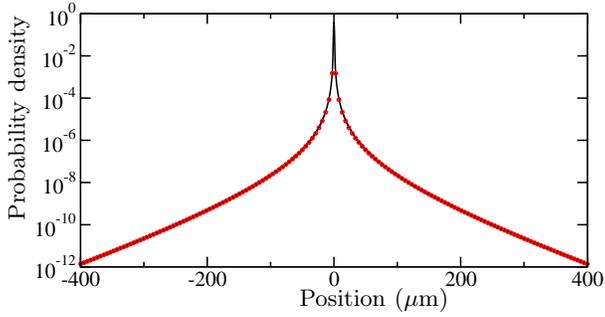}
\caption{Solid black line: stationary spatial probability density
\eref{pinfty} of a BEC wavepacket initially in the Thomas-Fermi regime
at disorder strength 
$\eta=0.2$ and $k_\text{max}\zeta = 0.65.$
Note the logarithmic scale and the fact that most of the wavepacket
remains well localized near the initial position, an effect
of the small-$k$-components of the wavepacket. Only in the far tails
where the density is extremely low  
one observes exponential localization with localization length
$\xiloc(k_\text{max})\approx 48\,\mu$m. 
The full shape of the wavepacket is well described
by $\exp[-r/\xiloc(k_\text{max})]/r^{5/2}$ (red circles).
}  
\label{fig:psi2}
\end{figure}

As noted in section \ref{ellB.sec} above, in 2D the transport mean-free path $\ellB(k)$ is
finite for all $k$-values.  Consequently, the localization length
$\xiloc(k)$ as given by \eref{xiloc} is also finite, and there is
exponential localization for all $k$-components in the initial
momentum distribution $\pi_0(\bk)$.  
This implies that a wavepacket will follow a diffusive behaviour at short time,
but the dynamics will slow down and finally freeze.
The resulting stationary probability distribution depends on the initial state.
Provided that the initial spatial size of the wavepacket
is small compared to the effective  width $\Delta r(k)\equiv \xiloc(k)$ of the propagation
kernel (for the $k$ values populated in the initial wavepacket),
the limit distribution reads 
\begin{equation}\label{pinfty}
\lim_{t \to \infty} p(r,t) \approx 
\mv{\frac{K_0(r/\xiloc(k))}{2\pi\xiloc^2(k)}
}_{\pi_0}. 
\end{equation}
This final density distribution is shown in fig.~\ref{fig:psi2} for the typical values
$\eta=0.2$ and $k_\text{max}\zeta = 0.65$ used throughout
this paper. Obviously, the shape near the center is not purely
exponential. Only quite far in the tails is the wavepacket component
at $k_\text{max}$ dominant, leading to 
exponential localization with localization length
$\xiloc(k_\text{max})\approx 48\,\mu$m.
Except very close to the origin, the density distribution is quite well described by 
$\exp[-r/\xiloc(k_\text{max})]/r^{5/2} $ as obtained by using the asymptotic behaviour of $K_0$ in eq.~(\ref{pinfty}). 

This behaviour in 2D can be compared to other dimensions. 
In 1D, and still within the Born approximation, two regimes have to be distinguished. In case the highest momentum lies still inside the support of the speckle two-point correlator, all $k$-components are exponentially localized, and an asymptotic behaviour $\exp(-|z|/\xiloc(k_\text{max}))/|z|^{7/2}$ is predicted \cite{Sanchez-Palencia2007}. In case the highest momenta are beyond the support of the two-point correlation, then a regime can be reached with an effective algebraic density decay as $|z|^{-2}$; these predictions have been recently confirmed by experiments \cite{Billy2008}. 
In 3D, averaging over the same momentum distribution leads to a $|r|^{-4}$ algebraic decay, without any exponential envelope. This is due to atoms with momenta close to the mobility edge, at least for the limiting case of uncorrelated Gaussian disorder \cite{Skipetrov2008}.
Thus, only in two dimensions do we have the favourable situation where
the density profile always shows an exponential localization. This is
because the two dimensional localization length is always finite for
all $k$-components for correlated disorder even within the Born
approximation.  

\begin{figure}
\psfrag{kzeta}{{$k\zeta$}}
\psfrag{xiloc}{{$\xiloc/\zeta$}}
\psfrag{dr}{{$\Delta r/2\zeta = \sqrt{\mv{\xiloc^2}_{\pi_0}}/\zeta$}}
\includegraphics[width=0.9\linewidth]{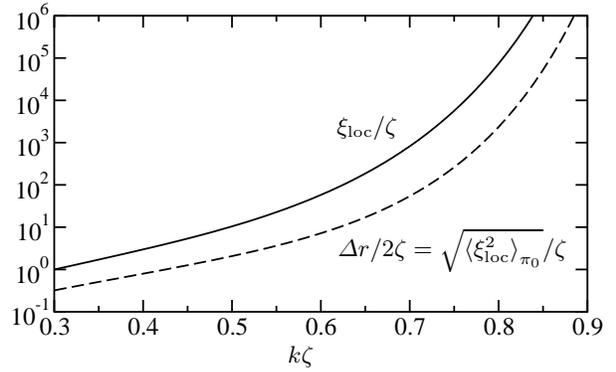}
\caption{Solid curve: localization length $\xiloc$ [eq.~\eref{xiloc}]
as a function of the atomic wave vector $k$
for $\eta=0.2$, 
both in units of the 2D-speckle
correlation length $\zeta$. 
Dashed line: stationary rms size  $\Delta
\br/2$ of the wavepacket [eq.~\eref{eq:varloca}] 
resulting from the expansion of an initial Thomas-Fermi wavepacket
with maximum momentum $k_\text{max}=k$.
As a result of the average, various shorter
localization lengths contribute  
and the resulting final size of the wavepacket is significantly
smaller than in the monochromatic case with the corresponding $k=k_\text{max}$.   
Note, however, that both curves display a similar rapid increase with $k$.}  
\label{fig:xiloc}
\end{figure}

Solving next for the variance \eref{variance} we get:
\begin{equation}\label{eq:varloca}
\lim_{t \to \infty} \Delta \br^2(t) = \Delta \br_0^2 +2d 
\mv{\xiloc^2(k)}_{\pi_0} 
\end{equation}
When the initial dispersion $\Delta \br_0^2 $ can be neglected, the
spatial variance will be given by 
the momentum-average of the squared localization length. 
This is the essence of strong localization: Transport starts in
the Boltzmann regime  and, as time increases, 
the diffusive dynamics gets slowed down by interference
effects. Finally the dynamics freezes in the stationary limit and    
2D transport always reaches the localized regime within a bulk system
in the absence of phase-breaking mechanisms. 
The time scale for the complete crossover from the diffusive to the stationary
localized distribution shown in Fig.~\ref{fig:psi2} is given by the 
localization time $\tauloc(k_\text{max})=\xiloc^2(k_\text{max})/\DB(k_\text{max}) $ for the fastest 
atoms with momentum $k_\text{max}$. Of course, some components of the
atomic cloud (the ones with lower momenta) will localize earlier and on a shorter
spatial scale, resulting in a strongly peaked probability density,
see Fig.~\ref{fig:psi2}.

\section{Quantum diffusion with limited phase coherence}
\label{wldephasing.sec}

Since 2D localization relies on interfering amplitudes over loops of arbitrary size   
and since the bulk localization length $\xiloc$ increases
exponentially fast with $k\ellB$, 
one must imperatively discuss spatial or temporal factors limiting
the interference effects.   
An important limitation is due to dephasing, and another 
obvious limitation is the finite size of the sample. Within the
framework established so far,  
we are able to take into account 
both limitations. First we 
consider a bulk system with dephasing, turning to limited system size in the following section \ref{size.sec}.

\subsection{Dephasing processes}

Perfect phase coherence is of course an idealization, and 
in real systems interference effects are killed by dephasing processes
at some rate $\gamma_\phi=\tau_\phi^{-1}$. This means that the
contribution of long diffusion loops to the interference-renormalized
diffusion constant is suppressed. 

To incorporate dephasing processes in the self-consis\-tent theory, we
implement the simple prescription  
$\omega \mapsto \omega + i/\tau_\phi$ in \eqref{returnproba}. This replacement 
can be shown to be exact within linear response theory for a rather large
class of different systems with microscopic dephasing sources, such as
electrons with  spin-flip scattering by magnetic impurites or photons
being scattered by atoms with a degenerate dipole transition
\cite{Mueller2005a}. Also mobile 
impurity atoms in a two-species experiment can be a source of
dephasing randomness \cite{Ospelkaus2006}, 
and in the limiting case of slowly moving impurities
an effective dephasing time is obtained that leads to the same qualitative
consequences as the above substitution \cite{Kogan2008}. 

It is important to stress here that the intensity propagation kernel \eqref{DiffusionKernel} keeps unchanged under dephasing, i.e., there $\omega$ is not to be shifted by $i/\tau_\phi$. This fact outlines an important difference between dephasing processes and absorption. Dephasing kills interference, hereby affecting the self-consistent diffusion constant given by \eqref{SCD}, but conserves particle number. Uniform absorption would cause particle losses at some constant rate $\gamma_a=\tau_a^{-1}$. It would not affect the relative phase between interfering amplitudes, but would damp their contribution as the loop size increases, thereby hindering the cumulative strength of interference effects.  
In the present paper, we consider conservative dynamics and therefore do not need to discuss absorption, which could be conveniently incorporated in the theory by the substitution $\omega \mapsto \omega + i/\tau_a$ both in \eqref{DiffusionKernel} and in \eqref{returnproba}. 

In the present context of atoms in optical potentials, we consider
that the main source of decoherence  
for the atom dynamics is the spontaneous emission of photons from the speckle field
 \cite{Kuhn2005,Kuhn2007}. These spontaneous emission processes occur at a rate 
\begin{equation} \label{tauphi} 
\frac{1}{\tauphi}= \frac{V \Gamma}{\hbar |\delta|}
\end{equation} 
where $\Gamma$ is the natural linewidth of the excited atomic state and 
$\delta$ is the detuning with respect to the atomic transition frequency. 

If the average time between successive spontaneous emission events is
of the order of---or even shorter than---the Boltzmann transport time,
$\tau_\phi\lesssim \tauB$, then this intense dephasing will be shown
below to result in purely classical transport already discussed in section
\ref{sec:boltzmann}.   
If on the contrary  $\tauphi\gg\tauB $, then the regime of coherent diffusive transport is reached. 
Since $\tau_\phi/\tauB$ scales as $V|\delta|/\Gamma$, this can be
always achieved at constant potential height provided the detuning
$|\delta|/\Gamma$ is sufficiently large.

\subsection{Weak localization  and residual diffusion}

After the replacement $\omega \to \omega + i/\tauphi$
in the self-consistent implicit
equation  \eref{Dlogfrequency}, we 
can now discuss its solution along the lines of section \ref{WSL.sec}.
Let us
consider
the stationary limit $\omega\to0$. The stationary diffusion constant solves
now the self-consistent equation   
\begin{equation} \label{Dsctauphi}
\frac{D(k)}{\DB(k)}=1- \frac{1}{\pi k\ellB}\log\left[
1+\frac{\tauphi D(k)}{\tauB \DB(k)}\right] . 
\end{equation} 

In the case of intense dephasing
$\tauphi\lesssim\tauB$, 
we get $D \to \DB$, a result easy to understand:
coherent effects are scrambled at such a high rate that 
Boltzmann diffusion 
is never affected by interference. The dynamics thus
stays entirely classical, and the results of 
section \ref{sec:boltzmann} apply. Of course, the
random momentum recoil transferred to the atoms by rapid spontaneous emission of
photons induces already an atomic
diffusion by itself \cite{Grynberg2001}, but here we assume that this effect is negligible
compared to the disorder-induced diffusion under study. 

More interesting is now the regime of  
phase-coherent transport $\tau_\phi\gg\tauB$. 
As in section \ref{WSL.sec}, the short-time dynamics $\omega\tauB\sim1$ 
is again dominated by the classical Boltzmann diffusion. 
However, 
the limit $\omega \to 0$ does not lead anymore to strong localization
in the strict sense. 
Instead, in the stationary limit, the residual dynamics stays diffusive. 
In the case of moderate dephasing, i.e.  
$\tau_\phi \ll \tauloc = \xiloc^2/\DB$, 
from \eref{Dsctauphi}
one finds the weak-localization correction  
to the diffusion constant in the form 
\begin{equation}\label{WLD}
\DWL(k) = \DB(k) - \frac{\hbar}{2\pi m} \ln \big( 1 + \frac{\tau_\phi}{\tauB} \big).
\end{equation} 

In the last case of very weak dephasing, 
$\tau_\phi \gg \tauloc$, the residual diffusion constant is 
\begin{equation} \label{ResD}
D(k) \approx \frac{\xiloc^2(k)}{\tau_\phi} = \DB(k)  \frac{\tauloc(k)}{\tau_\phi}. 
\end{equation} 
We can picture this regime by saying that coherent effects tend to localize the 
particle but phase-breaking mechanisms impose a residual slow
diffusion with  
$D \ll \DB$: the particle
propagates coherently over a relatively long time and thus localizes in time
$\tauloc$ on a scale $\xiloc$
until a phase-breaking event occurs (typically every $\tau_\phi$), resetting the diffusive
Boltzmann dynamics followed by localization after another $\tauloc,$ etc. 

\subsection{Localization onset in presence of dephasing}

\begin{figure}[t]
\centering
\psfrag{kzeta}{{$k\zeta$}}
\psfrag{DoB}{{$\dfrac{D}{\DB}$}}
\includegraphics[width=0.9\linewidth]{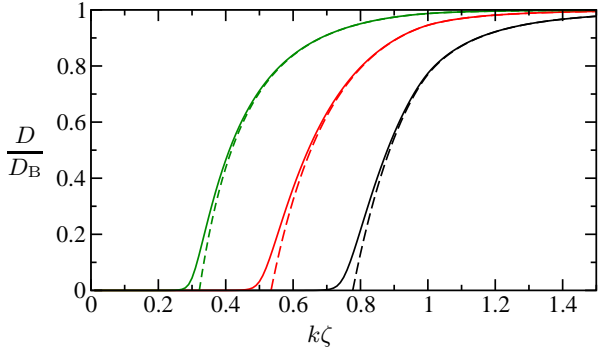}
\caption{Solid lines: diffusion constant $D/\DB$ given by eq.~\eref{Dsctauphi} as a
function of $k\zeta$ for increasing disorder strength (from left to right) 
$\eta=$0.05 (green), 0.1 (red), 
0.2 (black). 
Decoherence is driven here by spontaneous emission induced 
by the light-shift speckle field itself, cf.~\cite{Kuhn2005,Kuhn2007}. 
The speckle detuning is chosen at $\delta = 3.29\times 10^7 \Gamma$,
where $\Gamma$ is the width of the atomic excited state, realizing 
highly phase-coherent propagation with $\tauphi/\tauB\sim10^7$. 
The crossover from weak
localization at high $k$ to residual diffusion at
small $k$ is rather sharp. 
Dashed
curves: corresponding plots of the ratio $\DWL/\DB$ as given by
\eref{WLD}. The localization onset at $\kc$ where 
$\DWL =0$ gives a reasonably good approximation to the crossover.} 
\label{fig:D}
\end{figure}

Let us now examine more closely the crossover from weak to strong
localization as a function of relevant parameters. 
Figure \ref{fig:D} shows a plot of  $D/\DB$
solving \eref{Dsctauphi} (together with the
weak localization prediction \eref{WLD})
as a function of
$k\zeta$ for different values of disorder strength $\eta$. The laser
detuning from the atomic resonance is held fixed at $\delta =
3.29\times 10^7 \Gamma$. For $k\zeta<1$, the low-energy limit
\eref{lowk} then permits to estimate the effective dephasing rate as 
 $\tauphi/\tauB= 2\pi \eta |\delta|/\Gamma$, which varies slightly with
$\eta$ from curve to curve but is exceedingly large (of the order of
$10^7$) in all cases.  

Clearly, the curves show a sharp crossover 
between the weak localization regime at large momentum and residual
diffusion at small momentum.  
The crossover between these two regimes can be conveniently defined
through the condition $\tau_\phi = \tauloc$. 
Recasting the weak-localization corrections \eref{WLD} in the form 
\begin{equation}\label{WLC}
\frac{\DWL}{\DB} = \frac{1}{\pi  k\ellB} 
\ln\left(\frac{\tauB+\tauloc}{\tauB+\tau_\phi}\right),
\end{equation}
one immediately sees that the crossover condition $\tau_\phi =\tauloc$ is
reached at the point where the weakly localized diffusion constant
would vanish: $\DWL = 0$.  

\begin{figure}[t]
\psfrag{kc}{$\kc\zeta$}
\psfrag{eta}{{$\eta$}}
\centering
\includegraphics[width=0.9\linewidth]{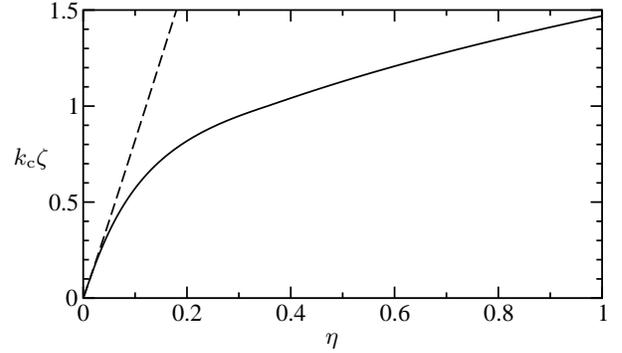}
\caption{Critical wavenumber $\kc\zeta$
(defining the effective onset of strong localization) as a function of the disorder strength
$\eta$ in the optical speckle field with correlation \eref{eq:cor2D} 
and detuning $\delta = 3.29\times 10^7 \Gamma$. 
 The dashed line is the low-energy behaviour \eref{kcdelta}. 
\label{fig:kc2d}}
\end{figure}

Solving for this condition defines an 
effective mobility edge $\kc$ in momentum space as a function of the disorder
parameters such as its strength $\eta=V/E_\zeta$.  
Figure \ref{fig:kc2d} shows a numerical plot of $\kc$ as a function of $\eta$ as obtained for the 
2D speckle correlator \eref{eq:cor2D} in the case of spontaneous
emission at detuning $\delta = 3.29\times 10^7 \Gamma$.
At low energy, one can use \eqref{lowk} to obtain:
\begin{equation}\label{kcdelta}
\kc\zeta = 2\eta \ \sqrt{\ln\left(1+\sfrac{2\pi\eta
|\delta|}{\Gamma}\right)},  \quad (k\zeta \ll 1). 
\end{equation} 
This low-energy asymptotics is also shown in figure
\ref{fig:kc2d}. Note that the weak-disorder condition $\eta \ll
\kc\zeta$ is well satisfied for $\eta \ll 1$. 

From a practical point of view, one can thus say 
that an effective strong localization regime is reached when only residual
diffusion with $D \ll \DB$ takes place. Consequently,  
one can earmark all momentum states with $k<\kc$ as ``localized'' 
and all states with $k>\kc$ as ``extended''. The localized fraction of atoms in this sense is 
\begin{equation}\label{eq:locfrac}
p_{\textrm{loc}}=\int_{k \leq \kc} \frac{\rmd\vec{k}}{(2\pi)^d} \ \pi_0(\vec{k}) = \beta^2(2-\beta^2)
\end{equation}
where $\beta = \min(1,\kc/k_\text{max})$. Not surprisingly, localizing
a large fraction of atoms 
 requires that almost all momenta present in the initial distribution are 
below the effective mobility edge.

\section{Finite-size systems}
\label{size.sec}

The impact of long diffusion loops is also limited by spatial boundaries 
in finite-size systems. 
As justified in \cite{Cherroret2008}, a correct way  to deal with 
finite-size systems is to introduce a \textit{space-dependent} 
renormalized diffusion constant. 
A detailed calculation also requires to know which boundary conditions
are imposed at the edges of the system. Obviously, periodic or fixed
boundary conditions (Dirichlet or Neumann) impose a discrete energy spectrum
while open boundary conditions give a continuous spectrum and
thus the impossibility of true 
localization (in the strict mathematical sense) at infinite time,
because of leakage at the edges.

A much simpler recipe is the replacement  $q^2 \to q^2 +1/L^2$ in the 
integrand of the return probability \eqref{SCD} (an equivalent alternative 
is to use $1/L$ as an infrared cutoff for the $q$-integral), thereby suppressing the 
contribution of long loops. 
Without a precise prescription of what happens at the boundaries (absorption, reflection with
a phase shift, periodic boundary conditions, etc.), this is the simplest approach. As we will see in the following,
it leads nevertheless to physically reasonable results in the localized regime. 
Of course, the system size must not be too small for diffusion to take
place, and one should have $L \gg \ellB$.   

In order to implement this simple approach, one may still 
work with the complex-frequency method used in the previous section 
and use the formal prescription 
$\omega \to \omega +i/\tauL$ with $\tauL = L^2/D$, a form which is suggestive of the 
average time it would take an 
atom with effective diffusion constant $D(\omega,k)$ to reach
the boundary. This cutoff time is obtained from the Thouless
time $\tau_\text{Th} =  L^2/\DB$ if one replaces the 
bare diffusion constant $\DB$ by the self-consistently
renormalized $D$. Note that although $\tauL = L^2/D$ formally acquires
an imaginary part if $D(\omega,k)$ does, the replacement in
\eqref{SCD} still provides a real momentum replacement $q^2 \to q^2
+1/L^2$ because  
$D(\omega,k)$ is divided out.

In principle, one should also use the (non-self-consistent) 
replacement $\omega \to \omega +i/\tau_\text{Th}$
in the intensity propagator \eqref{DiffusionKernel}. 
Like uniform absorption,  this leakage would lead to a global exponential
damping of the probability density, also known as the 
fundamental Holstein mode of the system \cite{SlowDiffusion}.
In the localized regime where the
system size is larger than the bulk localization length, $L\gg
\xiloc(k),$  
this damping (whose precise effect would again depend on the 
boundary conditions) is expected to be small and can be neglected.   

\subsection{Onset of localization in a finite-size system}

We now briefly discuss the case of a finite-size system in the 
absence of dephasing mechanisms ($\tau_\phi \to  \infty$). 
Using the finite-size replacement $\omega \to\omega+ i/\tau_L$ in the
self-consistent equation \eref{Dlogfrequency} one determines
the  
stationary diffusion constant in the limit $\omega\to 0$, which turns
out to be a non-zero real quantity when $\xiloc(k)>L:$  
\begin{equation}
D (k) = \frac{\hbar}{2\pi m} 
\ln\left[ \frac{\ellB^2(k) + \xiloc^2(k)}{\ellB^2(k) + L^2}\right]. 
\end{equation}
As this is precisely the regime where boundary conditions must be carefully specified,
this result must be taken with a grain of salt, and only gives an indication
of how much the classical transport is slowed down by interference effects.

The situation is more interesting below the critical momentum $\kc$
defined by the condition 
 $\xiloc(\kc) =L.$ There, the stationary diffusion constant vanishes, implying localization, 
and the asymptotic low-frequency behaviour is: 
\begin{equation}
\label{Domegaxi_prime}
D(\omega,k) \approx -i\omega \xiloc'^2(k). 
\end{equation}
with a modified localization length \cite{Kuhn2007}:
\begin{equation}
\label{xiloc_prime}
\frac{1}{\xiloc'^2(k)} = \frac{1}{\xiloc^2(k)} - \frac{1}{L^2}. 
\end{equation}
This expression says that cutting off very long loops 
in a finite system slightly increases the localization length. 
Note that this correction is algebraic
in $L/\xiloc$ while absorption at the boundary is expected
to be exponentially small. 
Les us repeat that this simple approach
is valid only if the system size is larger than the bulk localization length. 
When this condition is violated, a space-dependent 
diffusion constant and proper boundary conditions must be considered.

\subsection{Localization in a finite-size system with dephasing}

We account for the possibility of transport occuring inside a finite
domain while suffering from dephasing by summing the
respective rates, 
$\tau_0^{-1} = \tauL^{-1}+ \tauphi^{-1}$,  
a prescription reminiscent of the Matthiessen rule \cite{Ashcroft1976}. 
If one prefers to reason in terms of length scales, one may define 
the effective (complex) phase-coherence length $\ellphi= \sqrt{D\tauphi}$ and
use $L_0^{-1} = [L^{-2}+\ellphi^{-2}]^{1/2}$ as momentum cutoff through $q^2 \to q^2 + L_0^{-2}$. Since the complex diffusion constant is divided out, note again that this prescription is just equivalent to use real cutoffs through $-i \omega + D q^2 \to -i \omega + \tau_\phi^{-1} + D(q^2 + L^{-2})$.

\begin{figure}[t]
\centering
\psfrag{kzeta}{\figfont{$k\zeta$}}
\psfrag{DoD}{\figfont{$\dfrac{D}{\DB}$}}
\includegraphics[width=0.9\linewidth]{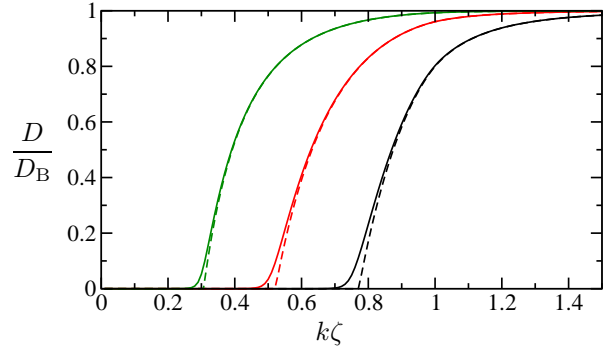}
\caption{Reduced diffusion constant $D(k)/\DB(k)$ as a
function of reduced atomic wave number $k\zeta$ for increasing disorder strength (from left to
right) $\eta=$0.05 (green), 0.1 (red), 
0.2 (black).
Dashed line: only the finite size $L=4\;$mm of the system is taken into
account. The onset of localization with vanishing  $D$ happens
when the localization length of the infinite system equals the
system size. 
Solid line: when dephasing effects such as spontaneous emission 
(detuning $\delta=3.29\times 10^7 \Gamma$) are taken into
account, the transition is smoothed.
}
\label{fs}
\end{figure}

If both finite size and dephasing are taken into
account, there is no strict onset for localization, but, as in the
situation with pure dephasing, an effective threshold
below which residual diffusion is tiny, $D \ll \DB.$
Figure~\ref{fs} shows the ratio of the diffusion constant---obtained by
solving the self-consistent equation
\eref{Dlogfrequency}---to the Boltzmann diffusion
constant,
when the finite size of the system is taken into account,
and when dephasing effects are included. Comparison with
fig.~\ref{fig:D} shows that the effect of the finite system
size in this case is rather small.

\section{Localization in experiments}
\label{exp.sec} 

\subsection{Effect of the initial momentum distribution}

Knowing the diffusion constant $D(k)$
for each momentum $k$, the average
diffusion constant \eref{averageD.eq} can be computed by integration
over the Thomas-Fermi momentum distribution 
\eqref{eq:pilaur} of an expanding BEC. 
One can use either the self-consistent diffusion constant
$D(k)$ or its approximation in the weak localization regime $\DWL(k).$
In the latter case, the integral must be cut below $\kc$ in order to avoid
counting an unphysical negative $\DWL$: 
\begin{equation}\label{WLAv}
\mathcal{D}_\text{WL} = \int_{k>\kc}\frac{\rmd\vec{k}}{(2\pi)^d} \, \pi_0(\vec{k}) \, \DWL(k). 
\end{equation}

\begin{figure}
\psfrag{DoB}{\figfont{$\dfrac{\mathcal{D}}{\boltz{\mathcal{D}}}$}}
\psfrag{kzeta}{\figfont{$k_\text{max}\zeta$}}
\centering
\includegraphics[width=0.9\linewidth]{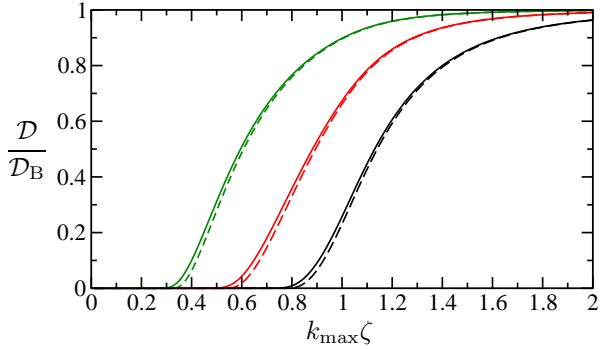}
\caption{Solid curves: ratio of momentum-averaged diffusion constants 
$\mathcal{D}/\boltz{\mathcal{D}}$ as a function of 
$k_\text{max}\zeta$ of an expanding BEC with momentum distribution \eref{eq:pilaur} 
for increasing disorder strength (from left to right) 
$\eta=$ 0.05 (green), 0.1 (red), 0.2 (black). 
The speckle correlation function is given by \eref{eq:cor2D}, 
the detuning has been fixed to $\delta = 3.29\times 10^7 \,
\Gamma$.  
Dashed curves: $\WL{\mathcal{D}}/\boltz{\mathcal{D}}$ obtained
from the weak-localization prescription \eref{WLD} and \eqref{WLAv}.
\label{fig:heal2d}}
\end{figure}

\fref{fig:heal2d} shows the ratio $\mathcal{D}/\boltz{\mathcal{D}}$ as
a function of $k_\text{max}\zeta$ for 
 different values of $\eta$ and the 2D speckle correlation \eref{eq:cor2D}. 
 A value of $\mathcal{D}/\boltz{\mathcal{D}}$ close to $1$ means that quantum corrections 
 to transport are difficult to observe. Conversely they become very strong when this value approaches zero. 
 Atoms are then practically completely localized. The threshold in $k$ to observe 
 localization increases with $\eta$, or equivalently, with $\kc$. We
 find again the criterion $k_\text{max}\sim \kc$ to observe ``strong
localization'' rather unambiguously. 

\begin{figure} 
\psfrag{DoB}{\figfont{$\dfrac{\mathcal{D}}{\mathcal{D}_B}$}}
\psfrag{ksigma}{\figfont{$\Delta k\zeta$}}
\centering
\includegraphics[width=0.9\linewidth]{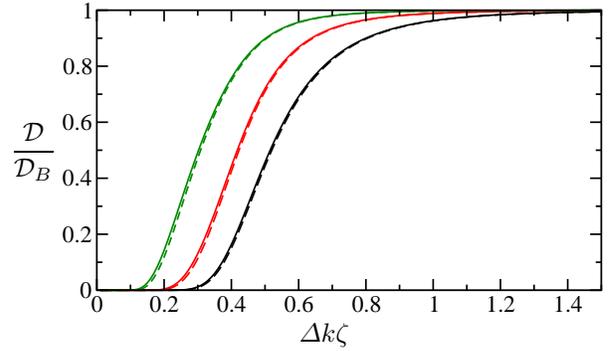}
\caption{Same as in figure \ref{fig:heal2d}, but for a Gaussian
momentum distribution with dispersion $\Delta k$. 
\label{fig:k0eq02d}}
\end{figure}

In \cite{Shapiro2007}, a Gaussian initial
wavepacket is discussed instead of a Thomas-Fermi distribution, 
leading to a Gaussian momentum distribution centered at $k=0$ with
dispersion $\Delta k$.  
\fref{fig:k0eq02d} gives our results for the ratio $\mathcal{D}/\mathcal{D}_B$ 
as a function of $\Delta k \zeta$ for different values of $\eta$. Compared to \fref{fig:heal2d}, 
there are no qualitative changes and no big quantitative changes. Hence the conclusions keep the same.

\subsection{Experimental realizability}

In order to estimate the orders of magnitude expected for a realistic
2D experiment, we will use the various experimental parameters of the
quasi-1D Orsay experiment \cite{Billy2008}. 
There, a BEC condensate of Rb$^{87}$ atoms ($m \approx 1.44 \, 10^{-25}\,$kg) 
is produced with a healing length $\xi \approx 0.40 \, \mu$m. The condensate
evolves  
in an optical speckle with  
correlation length $\zeta \approx 0.26\,\mu$m  
such that $k_{\mathrm{max}}\zeta = \zeta/\xi \approx 0.65$. The optical 
speckle  size $L \approx 4\,$mm 
is far-detuned from the Rb transition 
frequency by $\delta/\Gamma \approx 3.29\times10^7$.
The speckle fluctuation strength can be adjusted by varying the laser power,
up to $\eta \approx 0.046$. For reasons explained below, we performed calculations for 
$\eta=0.05, 0.1$ and 0.2. 
Since $\eta/k_{\mathrm{max}}\zeta$ is then at most 0.3 (and therefore sufficiently less than 1),
we can use the weak-disorder results, keeping in mind that exact numerical values
can differ from the ones computed in lowest-order
perturbation theory.  

Compared to the 1D situation, the main difficulty in 2D 
is to reach a sufficiently short transport mean free path $\ellB.$
Indeed, the exponential dependence of the localization length $\xiloc$
with 
$\ellB$---see eq.~(\ref{xiloc})---implies that $k\ellB$ must not be much larger than unity,
so that $k\xiloc$ itself can be at most of the order of few hundreds or thousands, meaning
$\xiloc$ smaller or comparable to 1 mm. Then, the Boltzmann diffusion constant, eq.~(\ref{DB}),
will be of the order of a few $\hbar/m$, that is a few $10^{-9}{\mathrm m}^2{\mathrm s}^{-1}.$
As the lifetime of an ultra-cold atomic gas is typically limited to a few seconds,
it is clear that the maximum diffusive expansion of the gas will be limited
to a few hundred $\mu$m. This in turn implies that localization by interference effects will be
observed only if the localization length is decreased to a sufficiently small value,
of the order of a fraction of a mm. This is very different from the 1D situation, where localization can
be observed even for a large transport mean free path (between $50\,\mu$m and $2\,$mm in~\cite{Billy2008})
and a large Boltzmann diffusion constant. 
A fortunate consequence is that the size of the speckle optical potential can be limited to a fraction
of mm, making it possible to focus the laser power on a smaller area, such that
values of $\eta$ up to 0.2 should be reachable.
  
With these parameter values, the effective
mobility edge is predicted to be at $\kc\zeta \approx$ 0.345, 0.571 and 0.818 for
$\eta=$ 0.05, 0.1 and 0.2, respectively. We thus predict that about $48\%, 95\%$ and $100\%$
of the atoms will be localized, respectively. 
Evidence of the 2D localized regime should thus be easily observable.

With $\eta=0.2,$ the maximum localization length (for atoms with $k=k_\text{max}$)
is about $48\,\mu$m, and the localization time around 2.6 s.
Loss of coherence by spontaneous emission shoud occur at a very small rate,
of the order of $\tau_{\phi}^{-1} \approx 6.7\times 10^{-5}\mathrm{s}^{-1}$,
leading to a tiny residual diffusion constant, of the order of $10^{-4}$ times the
Boltzmann diffusion constant. These very favorable numbers are due to the choice
of a far off-resonance laser beam to create the optical potential. A different
technical choice for the laser beam could of course lead to a different
situation.

In conclusion, these estimates show that strong localization of matter waves in 2D speckle potentials 
is within experimental reach. 


\section{Conclusion}

We have studied quantum diffusion of matter waves propagating in 2D optical speckle potentials within 
the framework of a self-consistent localization theory. We have 
consistently incorporated 
the effect of dephasing mechanisms 
and system boundaries. We have calculated the average probability density
$p(\br,t)$ of matter waves propagating in a disordered potential for
any initial Wigner phase-space distribution. This 
quantity allows to calculate the variance of the expanding cloud. 
We have shown that dephasing processes prevent 
true localization from taking place and induce a residual diffusion in the long-time limit. 
In certain regimes of parameters, the resulting stationary diffusion constant is 
so small that, for all practical purposes, one can define a
localization onset by the momentum where the bulk weakly localized
diffusion constant vanishes. This mobility edge 
then separates ``localized'' states from ``extended'' ones, 
allowing for a practical definition of the localized fraction of atoms. 
These quantities serve as 
experimental parameters to study weak and strong localization from a practical 
point of view. 

We have specialized our results to the case of an inter\-action-driven
BEC expanding in the presence of the disordered optical potential and to the case 
of a Gaussian momentum distribution. In any case strong localization can only be 
achieved for cold enough atoms. 
Taking state-of-the-art figures 
we have shown that Anderson localization of matter waves in 
2D optical speckle potentials should be observable in near-future experiments.

\begin{acknowledgement}
The authors would like to thank Guillaume Labeyrie for raising this subject
and for many helpful discussions about this work.
This work was supported by the DAAD, the
BFHZ-CCUFP and the Marie Curie program (contract number
HPMT-2000-00102) at Laboratoire Kastler Brossel (Universit\'e
Pierre et Marie Curie / \'Ecole Normale Sup\'erieure, UMR 8552 du
CNRS). RK would like to thank Professor B.-G. Englert and the
Quantum Information Lab at the National University of Singapore
for their kind hospitality (A*STAR Temasek Grant 012-104-0040).
\end{acknowledgement}


\begin{thebibliography}{31}

\bibitem{Lewenstein2007}
M.~Lewenstein, A.~Sanpera, V.~Ahufinger, B.~Damski, A.~Sen, U.~Sen, Adv. Phys.
  \textbf{56}, 243 (2007)

\bibitem{Anderson1958}
P.W. Anderson, Phys. Rev. Lett. \textbf{109}, 1492 (1958)

\bibitem{Tiggelen1999}
B.v. Tiggelen, in \emph{Diffuse Waves in Complex Media}, edited by J.P. Fouque
  (Kluwer, 1999), pp. 1--60

\bibitem{Kramer1993}
B.~Kramer, A.~MacKinnon, Rep. Progr. Phys. \textbf{56}, 1469 (1993)

\bibitem{Billy2008}
J.~Billy, V.~Josse, Z.~Zuo, A.~Bernard, B.~Hambrecht, P.~Lugan, D.~Cl\'ement,
  L.~Sanchez-Palencia, P.~Bouyer, A.~Aspect, Nature \textbf{453}, 891
  (2008)

\bibitem{Roati2008}
G.~Roati, C.~D'Errico, L.~Fallani, M.~Fattori, C.~Fort, M.~Zaccanti,
  G.~Modugno, M.~Modugno, M.~Inguscio, Nature \textbf{453}, 895 (2008)

\bibitem{Abrahams1979}
E.~Abrahams, P.W. Anderson, D.C. Licciardello, T.V. Ramakrishnan, Phys. Rev.
  Lett. \textbf{42} (1979)

\bibitem{Kuhn2005}
R.C. Kuhn, C.~Miniatura, D.~Delande, O.~Sigwarth, C.A. M{\"u}ller, Phys. Rev.
  Lett. \textbf{95}, 250403 (2005)

\bibitem{Kuhn2007}
R.~Kuhn, O.~Sigwarth, C.~Miniatura, D.~Delande, C.A. M{\"u}ller, New J. Phys.
  \textbf{9}, 161 (2007)

\bibitem{Vollhardt1980}
D.~Vollhardt, P.~W{\"o}lfle, Phys. Rev. B \textbf{22}, 4666 (1980)

\bibitem{Shapiro1982} B.~Shapiro, Phys. Rev. B \textbf{25}, 4266 (1982)

\bibitem{Sanchez-Palencia2007}
L.~Sanchez-Palencia, D.~Cl\'ement, P.~Lugan, P.~Bouyer, G.V. Shlyapnikov,
  A.~Aspect, Phys. Rev. Lett. \textbf{98}, 210401 (2007)

\bibitem{Shapiro2007}
B.~Shapiro, Phys. Rev. Lett. \textbf{99}, 060602 (2007)

\bibitem{Clement2006}
D.~Cl\'ement, A.F. Var\'on, J.A. Retter, L.~Sanchez-Palencia, A.~Aspect,
  P.~Bouyer, New J. Phys. \textbf{8}, 165 (2006)

\bibitem{Goodman1975}
J.W. Goodman, in \emph{{Laser speckle and related phenomena}}, edited by J.C.
  Dainty (Springer-Verlag, 1975)

\bibitem{Hartung2008}
M.~Hartung, T.~Wellens, C.A. M{\"u}ller, K.~Richter, P.~Schlagheck, Phys. Rev.
  Lett. \textbf{101}, 020603 (2008)


\bibitem{Hillery1984}
M.~Hillery, R.F. O'Connell, M.O. Scully, E.P. Wigner, Phys. Rep.
  \textbf{106}, 121 (1984)

\bibitem{Rammer1998}
J.~Rammer, \emph{{Quantum Transport Theory}} (Perseus Books, Reading, Mass.,
  1998)

\bibitem{Kagan1996}
Y.~Kagan, E.L. Surkov, G.V. Shlyapnikov, Phys. Rev. A \textbf{54}, R1753 (1996)

\bibitem{Ashcroft1976}
N.~Ashcroft, D.~Mermin, \emph{{Solid State Physics}} (Saunders College,
  Philadelphia, 1976)

\bibitem{Akkermans2007}
E.~Akkermans, G.~Montambaux, \emph{Mesoscopic physics of electrons and photons}
  (Cambridge University Press, Cambridge, 2007)

\bibitem{Chandrasekhar1960}
S.~Chandrasekhar, \emph{{Radiative Transfer}} (Dover Publications, New York,
  1960)
  
 \bibitem{Apalkov2004} V.~M.~Apalkov, M.~E.~Raikh and B.~Shapiro, J. Opt. Soc. Am. B \textbf{21}, 132 (2004)

\bibitem{Akkermans1995}
E.~Akkermans, G.~Montambaux, J.L. Pichard, J.~Zinn-Justin, \emph{Mesoscopic
  Quantum Physics (Proceedings of the Les Houches Summer School Session LX1)}
  (Elsevier Science, Amsterdam, 1995)

\bibitem{Lobkis2005}
O.I. Lobkis, R.L. Weaver, Phys. Rev. E \textbf{71}, 011112 (2005)

\bibitem{Gorkov1979}
L.~Gor'kov, A.~Larkin, D.~Khmel'nitski\u{\i}, Pis'ma Zh. Eksp. Teor. Fiz.
  \textbf{30}, 248 (1979)  [JETP Lett. \textbf{30}, 228 (1979)]

\bibitem{Bergmann1984}
G.~Bergmann, Phys. Rep. \textbf{107}, 1 (1984)

\bibitem{Abramowitz1972}
M.~Abramowitz, I.~Stegun, \emph{{Handbook of mathematical functions with
  formulas, graphs, and mathematical table}} (Dover Publications, New York,
  1972)

\bibitem{Skipetrov2008}
S. E. Skipetrov, A. Minguzzi, B. A. van Tiggelen, B. Shapiro, Phys. Rev. Lett.
	\textbf{100}, {165301} (2008)

\bibitem{Mueller2005a}
C.A. M{\"u}ller, C.~Miniatura, E.~Akkermans, G.~Montambaux, J. Phys. A: Math.
  Gen. \textbf{38}, 7807 (2005)

\bibitem{Ospelkaus2006}
S.~Ospelkaus, C.~Ospelkaus, O.~Wille, M.~Succo, P.~Ernst, K.~Sengstock,
  K.~Bongs, Phys. Rev. Lett. \textbf{96}, 180403 (2006)

\bibitem{Kogan2008}
E.~Kogan, Eur. Phys. J. B \textbf{61}, 181 (2008)

\bibitem{Grynberg2001}
G.~Grynberg, C.~Robilliard, Phys. Rep. \textbf{355}, 335 (2001)

\bibitem{Cherroret2008}
N.~Cherroret, S.E. Skipetrov, Phys. Rev. E \textbf{77}, 046608 (2008)

\bibitem{SlowDiffusion}
G.~Labeyrie, E.~Vaujour, C.A.~M\"uller, D.~Delande, C.~Miniatura, D.~Wilkowski, R.~Kaiser
Phys. Rev. Lett. \textbf{91}(22), 223904 (2003)

\end{thebibliography}
\end{document}